\documentclass[twocolumn,english,pre]{revtex4-1}
\usepackage[T1]{fontenc}
\usepackage[latin9]{inputenc}
\usepackage{color}
\usepackage{amsmath}
\usepackage{graphicx}

\makeatletter
\usepackage{babel}

\makeatother

\usepackage{babel}
\begin{document}
\title{Thermodynamics of stationary states of the ideal gas in a heat flow}
\author{Robert Ho\l yst}
\email{equal contribution; rholyst@ichf.edu.pl}

\author{Karol Makuch}
\email{equal contribution; kmakuch@ichf.edu.pl}

\affiliation{Institute of Physical Chemistry, Polish Academy of Sciences Kasprzaka
44/52, 01-224 Warszawa}
\author{Anna Macio\l ek}
\affiliation{Institute of Physical Chemistry, Polish Academy of Sciences Kasprzaka
44/52, 01-224 Warszawa}
\affiliation{Max-Planck-Institut f{ü}r Intelligente Systeme Stuttgart, Heisenbergstr.~3,
D-70569 Stuttgart, Germany}
\author{Pawe\l{} J. \.{Z}uk}
\affiliation{Institute of Physical Chemistry, Polish Academy of Sciences Kasprzaka
44/52, 01-224 Warszawa}
\affiliation{Department of Physics, Lancaster University, Lancaster LA1 4YB, United
Kingdom}
\begin{abstract}
There is a long-standing question as to whether and to what extent
it is possible to describe nonequilibrium systems in stationary states
in terms of global thermodynamic functions. The positive answers have
been obtained only for isothermal systems or systems with small temperature
differences. We formulate thermodynamics of the stationary states
of the ideal gas subjected to heat flow in the form of the zeroth,
first, and second law. Surprisingly, the formal structure of steady
state thermodynamics is the same as in equilibrium thermodynamics.
We rigorously show that $U$ satisfies the following equation $dU=T^{*}dS^{*}-pdV$
for a constant number of particles, irrespective of the shape of the
container, boundary conditions, size of the system, or mode of heat
transfer into the system. We calculate $S^{*}$ and $T^{*}$ explicitly.
The theory selects stable nonequilibrium steady states in a multistable
system of ideal gas subjected to volumetric heating. It reduces to
equilibrium thermodynamics when heat flux goes to zero. 
\end{abstract}
\maketitle

\section{Introduction}

Thermodynamics simplifies the description of equilibrium systems.
It reduces the number of equations of state of material by expressing
them with one formula in terms of a thermodynamic potential \citep{Thermodynamics_and_an_Introduction_to_Thermostatistics_2ed_H_Callen}.
This simplification also significantly reduces the number of measurements
needed to determine any material's equilibrium properties \citep{History_of_Thermodynamics_The_Doctrine_of_Energy_and_Entropy_by_Ingo_Muller}.
It also determines the equilibrium state of the system by optimization
rules.

For similar reasons, there has been an enormous research effort to
introduce global thermodynamics with optimization rules and potential-like
formulation for steady states \citep{Introduction_to_thermodynamics_of_irreversible_processes_Ilya_Prigogine,oono1998steady,sekimoto1998langevin,hatano2001steady,sasa2006steady,guarnieri2020non,boksenbojm2011heat,netz2020approach,mandal2013nonequilibrium,holyst2019flux,jona2014thermodynamics,speck2005integral,mandal2016analysis,maes2019nonequilibrium,Zhang2021continuous,glansdorff1964general,maes2014nonequilibrium,ruelle2003extending,sasa2014possible,komatsu2008expression,komatsu2008steady,komatsu2011entropy,chiba2016numerical,nakagawa2017liquid,nakagawa2019global,sasa2021stochastic,nakagawa2022unique}.
The progress in this direction is limited either to the isothermal
situations or to the small temperature differences \citep{komatsu2008expression,komatsu2008steady,komatsu2011entropy,chiba2016numerical,nakagawa2017liquid,nakagawa2019global,sasa2021stochastic,nakagawa2022unique}.
Here we break this limitation and show that the steady state thermodynamic
description also exists for a system that is far from equilibrium
(with large temperature gradients).

\begin{figure}[bp]
\includegraphics[width=8.5cm]{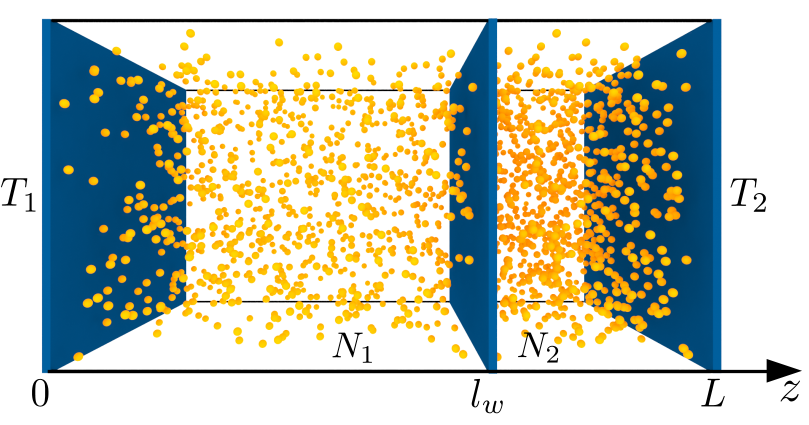}\caption{\textcolor{black}{Schematic illustration of an ideal gas in a box
with an internal wall.}}
\end{figure}

For a paradigmatic system such as an ideal gas, only three parameters
(entropy $S$, volume $V$, and the number of particles $N$) are
sufficient to determine its state at thermal equilibrium. At a nonequilibrium
state, one has to consider spatially dependent parameters, such as
temperature $T\left(\mathbf{r}\right)$, pressure $p\left(\mathbf{r}\right)$,
and density $n\left(\mathbf{r}\right)$. A description of such fields
appears in De Groot and Mazur's monograph on irreversible thermodynamics
\citep{Groot_Mazur_Non-equilibrium_thermodynamics}. This description
is based on local conservation laws of mass, momentum and energy.
With the assumption of local equilibrium and constitutive relations,
irreversible thermodynamics determines the state of the system.

In this paper, we show that within irreversible thermodynamics, there
exists a global description of a steady state of an ideal gas in heat
flow. Within this description the total system's energy is represented
as a function of $S^{*}$, volume $V$, and the number of particles
$N$. We formulate thermodynamics of steady state in the form of the
zeroth, first and second law and determine $S^{*}$ explicitly. Our
scheme for thermodynamics of nonequilibrium steady states is rigorous
and valid for large heat flux.

We illustrate the scheme using a monoatomic ideal gas confined between
two parallel walls with different temperatures $T_{1}$ and $T_{2}$.
Furthermore, we introduce to this system a constraint in the form
of a thin wall separating the gas into two parts, as shown in Fig.
1. We assume that this internal wall is diathermic and impenetrable.
Considering the system with internal constraints puts our problem
in the perspective of equilibrium thermodynamics as described by Callen
 `The single, all-encompassing problem of thermodynamics is the determination
of the equilibrium state that eventually results after the removal
of internal constraints in a closed, composite system' \citep{Thermodynamics_and_an_Introduction_to_Thermostatistics_2ed_H_Callen}.
We show the minimum principle that determines the stable position
of the internal wall.

This description also applies to different shapes of the system, boundary
conditions, and different modes of heat transfer (heat flows through
the system or heat is generated inside the system) and is valid beyond
the regime of linear irreversible thermodynamics, thus taking into
account the temperature-dependent heat conductivity.

\section{Ideal gas in heat flow}

We consider a fluid described by irreversible thermodynamics \citep{Groot_Mazur_Non-equilibrium_thermodynamics}.
Therefore, the gas is described by five equations: two local equations
of state and three conservation laws (continuity equation, Navier-Stokes
equation, and the energy balance equation) supplemented by proper
boundary conditions. We assume that the equation of state corresponds
to the monoatomic ideal gas. The gas is confined between two parallel
walls at positions $0$ and $L$. We assume that the system is translationally
invariant in $x,y$ directions. Moreover, the gas satisfies local
equilibrium and is described by the following equations of state \citep{Thermodynamics_and_an_Introduction_to_Thermostatistics_2ed_H_Callen}
\begin{equation}
p\left(z\right)=n\left(z\right)k_{B}T\left(z\right),\label{eq:eq of state pressure}
\end{equation}
with Boltzmann constant $k_{B}$, pressure $p\left(z\right)$, particle
number density $n\left(z\right)$, and the temperature $T\left(z\right)$
at position $z$. It is worth mentioning that the local equilibrium
is sometimes questioned. But as we discuss in the conclusions section,
for an ideal gas the local equilibrium is valid up to extreme temperature
gradients of the order of $10^{7}K/cm$. The energy equation of state
is given by 
\begin{equation}
\epsilon\left(z\right)=\frac{3}{2}n\left(z\right)k_{B}T\left(z\right).\label{eq:energy eos}
\end{equation}
Here $\epsilon\left(z\right)$ is the internal energy volumetric density.
We also assume that in the whole volume $V=AL$, where $A$ is an
area (along $x,y$) direction, there are $N$ particles. The boundary
condition follows from the assumption of a given temperature on the
walls 
\begin{align}
T\left(0\right) & =T_{1},\nonumber \\
T\left(L\right) & =T_{2}.\label{eq:temp bc external walls}
\end{align}

In the stationary state, the system is described by $T_{1},T_{2},A,L,N$,
which we call the control parameters. We assume that there is no mass
flow for the confined gas in a stationary state. That simplifies the
thermohydrodynamic equations \citep{Groot_Mazur_Non-equilibrium_thermodynamics}.
The Navier-Stokes equation is reduced to a condition of vanishing
pressure gradient, $\nabla p\left(z\right)=0.$ The two equations
of state follow that the energy density is also constant in space
and $\epsilon=3p/2$. The total energy $U$ is thus given by $U=A\int_{0}^{L}dz\,\epsilon=3ALp/2$.
We rewrite this expression in terms of the volume of the system, $V$
, obtaining the relation between pressure and volume, 
\begin{equation}
p=\frac{2}{3}\frac{U}{V}.\label{eq:pressure U and V}
\end{equation}

The energy balance equation with the Fourier law for the heat flux,
\begin{equation}
J_{q}=-\kappa\nabla T\left(z\right),\label{eq:fourier law}
\end{equation}
gives $0=\kappa\frac{d^{2}}{dz^{2}}T\left(z\right).$ With the boundary
conditions (\ref{eq:temp bc external walls}), it yields a linear
temperature profile, $T\left(z\right)=T_{1}+\left(T_{2}-T_{1}\right)z/L$.
With the constant pressure and the equation of state, it determines
density profile, $n\left(z\right)=p/k_{B}T\left(z\right)$, and with
a given number of particles, $N=A\int_{0}^{L}dz\,n\left(z\right)$,
they determine pressure, 
\begin{equation}
p=\frac{N}{V}k_{B}\frac{T_{2}-T_{1}}{\log\frac{T_{2}}{T_{1}}}.\label{eq:p by T1 T2}
\end{equation}

Suppose that the system is in the stationary state described by parameters
$T_{1},T_{2},A,L,N$. Then we start changing the temperature $T_{2}$
to $T_{2}+dT_{2}$. After a while the system reaches another stationary
state, this time described by parameters $T_{1},T_{2}+dT_{2},A,L,N$.
We could similarly move one of the system's walls and change its length
$L\to L+dL$. The disturbance of the system induces time-dependent
thermo-hydrodynamic flows. They may be complex (with sound waves,
turbulent motion or heat front \cite{zuk2022transient}) if the change
of the temperature or position of the wall is sudden. Nevertheless,
the possibility of solving thermohydrodynamic equations would allow
one to monitor the change of the internal energy $dU$, the net heat
$\mkern3mu \mathchar'26\mkern-12mu dQ$ entering the system during
the transition, and the work done $\mkern3mu \mathchar'26\mkern-12mu dW$.
Independently of the rate of change of the control parameters, the
energy balance within irreversible thermodynamics must have the following
consequences in the context of passing from one to another stationary
state, 
\begin{equation}
dU=\mkern3mu \mathchar'26\mkern-12mu dW+\mkern3mu \mathchar'26\mkern-12mu dQ.\label{eq:energy balance dU}
\end{equation}
The energy change is determined by the control parameters through
equations (\ref{eq:pressure U and V}) and (\ref{eq:p by T1 T2}).
As in similar considerations within equilibrium thermodynamics, the
work and heat of the transitions between steady states depend on the
transition rate. However, there is an essential simplification for
the case of slow processes \citep{oono1998steady}. We expect, that
the slow change of the boundary condition does not disturb the homogeneity
of the pressure in the system. In the limit of slow changes, the work
done in the transition is given by 
\begin{align}
\mkern3mu \mathchar'26\mkern-12mu dW & =-pdV,\label{eq:definition of work}
\end{align}
where $dV$ denotes the differential of the volume of the system.
Therefore the net heat differential is determined from (\ref{eq:energy balance dU})
and (\ref{eq:definition of work}) by 
\begin{equation}
\mkern3mu \mathchar'26\mkern-12mu dQ=dU+pdV.\label{eq:ex heat differential 2}
\end{equation}
We prove it below using thermo-hydrodynamic equations. The above equation
represents the energy balance in the system. It may be called the
first law because it has a form of and it reduces to the first law
of equilibrium thermodynamics when the heat does not flow through
the gas.

\section{The first law for stationary states in the case of thermohydrodynamics}

The total internal energy of the gas at given time instant, $U_{i}\left(t\right)$,
in volume $V\left(t\right)$ is given by the following integration
of its density, 
\begin{equation}
U_{i}\left(t\right)=\int_{V(t)}d^{3}r\,\rho\left({\bf r},t\right)u\left({\bf r},t\right),\label{eq:internal energy}
\end{equation}
where $\rho\left({\bf r},t\right)$ is the density and $u\left({\bf r},t\right)$
is the internal energy density per unit mass at position ${\bf r}$
and time $t$. To facilitate our considerations, but without loss
of generality, we assume the translational invariance of the system
in $x,y$ directions. Before time $t_{i}$ the system is in a stationary
state. Then, due to a change of volume $V$, temperatures $T_{1}$,
$T_{2}$, or other external factors, the system is taken to another
nearby stationary state, which is achieved after $t_{f}$. For example,
between times $t_{i}$ and $t_{f}$ we slowly change the position
of the right wall by manipulating its position, $L\left(t\right)$,
such that initially $L\left(t_{i}\right)=L$ changes to $L\left(t_{f}\right)=L+dL$.
That gives the time dependent volume, $V\left(t\right)=AL\left(t\right)$,
with the total change $dV=AdL$ when passing from the stationary state
at $t_{i}$ to the stationary state at $t_{f}$. Differential of the
energy (i.e. small change of the energy when passing to neighboring
stationary state) is given by, 
\[
dU=U_{i}\left(t_{f}\right)-U_{i}\left(t_{i}\right),
\]
which we equivalently express as, 
\begin{equation}
dU=\int_{t_{i}}^{t_{f}}dt\,\frac{dU_{i}\left(t\right)}{dt}.\label{eq:u differential}
\end{equation}
With the use of Eq. (\ref{eq:internal energy}) we get $dU_{i}/dt=\frac{d}{dt}\int_{V(t)}d^{3}r\,\rho\left({\bf r},t\right)u\left({\bf r},t\right)$.
In this expression, the integral is simplified with the use of $x,y$
translational symmetry and keeping in mind, that $V\left(t\right)=AL\left(t\right)$
as follows, 
\[
\int_{V(t)}d^{3}r\,\rho\left({\bf r},t\right)u\left({\bf r},t\right)=A\int_{0}^{L\left(t\right)}dz\,\rho\left(z,t\right)u\left(z,t\right),
\]
so the time derivative of internal energy reduces to 
\begin{align}
\frac{dU_{i}}{dt} & =A\frac{d}{dt}\int_{0}^{L\left(t\right)}dz\,\rho\left(z,t\right)u\left(z,t\right)\nonumber \\
 & =A\rho\left(L\left(t\right),t\right)u\left(L\left(t\right),t\right)\frac{dL\left(t\right)}{dt}\nonumber \\
 & +A\int_{0}^{L\left(t\right)}dz\,\frac{\partial}{\partial t}\left[\rho\left(z,t\right)u\left(z,t\right)\right].\label{eq:dU help}
\end{align}
In the latter integral there appears the left-hand side of the balance
energy equation (cf. p. 18 in \citep{Groot_Mazur_Non-equilibrium_thermodynamics})
\begin{align}
\frac{\partial}{\partial t}\left[\rho\left(z,t\right)u\left(z,t\right)\right] & =-div\left(\rho\left(z,t\right)u\left(z,t\right){\bf v}\left(z,t\right)+{\bf J}_{q}\right),\nonumber \\
 & -{\bf P}:grad{\bf v}\left(z,t\right)\label{eq:energy balance}
\end{align}
with velocity field ${\bf v}\left(z,t\right)$, heat flow $\boldsymbol{J}_{q}$
and pressure tensor ${\bf P}=p\left(z,t\right){\bf I}+$${\bf \Pi}$,
where ${\bf I}$ is the unit 3-dimensional matrix and ${\bf \Pi}$
is proportional to velocity gradients. Due to the fact, that there
is no velocity field in the system in stationary state, and the change
of the parameters is slow, we keep only the leading terms in velocity
field in the above expression, neglecting the quadratic term, ${\bf \Pi}:grad{\bf v}\left(z,t\right)\approx0$.
Therefore ${\bf P}:grad{\bf v}\left(z,t\right)\approx p\left(z,t\right)\,div{\bf v}\left(z,t\right)$
and the energy balance equation simplifies to 
\begin{align*}
 & \frac{\partial}{\partial t}\left[\rho\left(z,t\right)u\left(z,t\right)\right]=\\
 & -div\left(\rho\left(z,t\right)u\left(z,t\right){\bf v}\left(z,t\right)+{\bf J}_{q}\right)-p\left(z,t\right)\,div{\bf v}\left(z,t\right).
\end{align*}
Using the above in expression (\ref{eq:dU help}) we obtain 
\begin{align*}
\frac{dU_{i}}{dt} & =A\rho\left(L\left(t\right),t\right)u\left(L\left(t\right),t\right)\frac{dL(t)}{dt}\\
 & -A\int_{0}^{L\left(t\right)}dz\,div\left(\rho\left(z,t\right)u\left(z,t\right){\bf v}\left(z,t\right)\right)+\\
 & -A\int_{0}^{L\left(t\right)}dz\,div{\bf J}_{q}-A\int_{0}^{L\left(t\right)}dz\,p\left(z,t\right)\,div{\bf v}\left(z,t\right).
\end{align*}
The first two terms on the right-hand side give zero, because 
\begin{align*}
 & \int_{0}^{L\left(t\right)}dz\,div\left(\rho\left(z,t\right)u\left(z,t\right){\bf v}\left(z,t\right)\right)=\\
 & \rho\left(L\left(t\right),t\right)u\left(L\left(t\right),t\right){\bf v}_{z}\left(L\left(t\right),t\right)-\rho\left(0,t\right)u\left(0,t\right){\bf v}_{z}\left(0,t\right)
\end{align*}
and because $z$-component of the velocity above vanishes for $z=0$
and is equal to $dL\left(t\right)/dt$ for $z=L$. The third term,
$A\int_{0}^{L\left(t\right)}dz\,div{\bf J}_{q}=\int_{V\left(t\right)}d^{3}r\,div{\bf J}_{q}$
is the total heat rate that flows to the system which is evident after
application of Gauss theorem, $\int_{V\left(t\right)}d^{3}r\,div{\bf J}_{q}=\int_{\partial V\left(t\right)}d^{2}r\,{\bf n}\cdot{\bf J}_{q}$.
Here, ${\bf n}$ is the normal vector pointing outside the surface.
We denote the heat rate flowing into the system by 
\begin{equation}
q\left(t\right)\equiv-\int_{\partial V\left(t\right)}d^{2}r\,{\bf n}\cdot{\bf J}_{q}.\label{eq:heat rate}
\end{equation}
To simplify the fourth term we use the fact, that pressure in the
system during slow change of parameters is still homogeneous, $p\left(z,t\right)=p\left(t\right)$,
therefore $\int_{0}^{L\left(t\right)}dz\,p\left(z,t\right)\,div{\bf v}\left(z,t\right)=Ap\left(t\right)\int_{0}^{L\left(t\right)}dz\,\,div{\bf v}\left(z,t\right)$.
This integral is the volume change rate, $dV\left(t\right)/dt=\int_{V\left(t\right)}d^{3}r\,\,div{\bf v}\left(z,t\right)$
and finally for the fourth term we get, $A\int_{0}^{L\left(t\right)}dz\,p\left(z,t\right)\,div{\bf v}\left(z,t\right)=p\left(t\right)dV\left(t\right)/dt.$
Therefore the change of the energy (\ref{eq:dU help}) simplifies
to 
\[
\frac{dU_{i}}{dt}=q\left(t\right)-p\left(t\right)\frac{dV\left(t\right)}{dt}.
\]
Utilizing the above in expression (\ref{eq:u differential}) we obtain,
\begin{equation}
dU=\int_{t_{i}}^{t_{f}}dt\,q\left(t\right)-\int_{t_{i}}^{t_{f}}dt\,p\left(t\right)\frac{dV\left(t\right)}{dt}.\label{eq:du by time integral}
\end{equation}
We use the dominant term for small changes of parameters (neighboring
stationary state) in $\int_{t_{i}}^{t_{f}}dt\,p\left(t\right)\frac{dV\left(t\right)}{dt}\approx p\int_{t_{i}}^{t_{f}}dt\,\frac{dV\left(t\right)}{dt}=pdV.$
The above energy differential may be written in the form of (\ref{eq:energy balance dU})
where 
\[
\mkern3mu \mathchar'26\mkern-12mu dQ=\int_{t_{i}}^{t_{f}}dt\,q\left(t\right)
\]
is the total heat transfer to the system and $\mkern3mu \mathchar'26\mkern-12mu dW=-pdV$
is the work performed on the system during the transition between
stationary states.

It is worth noting that in the above derivation, we did not specify
temperature changes. Therefore the energy balance (\ref{eq:energy balance dU})
is valid for transitions in the space of $V,T_{1},T_{2}$. Eq. (\ref{eq:energy balance dU})
derived above is valid under the assumption of slow changes of external
parameters (including homogeneous pressure condition). In this limit
it is a rigorous expression. Therefore, if $q\left(t\right)=0$ in
a stationary state (before $t_{i}$ and after $t_{f}$) and both $dU$,
$\mkern3mu \mathchar'26\mkern-12mu dW$ are finite and well defined
(which is exactly the case considered here), then the net heat, $\mkern3mu \mathchar'26\mkern-12mu dQ$,
transferred to the system during the transition is finite and well
defined as well.

\section{nonequilibrium entropy as a potential of the net heat differential}

It is worth noting that the net heat introduced above would be the
excess heat considered by Oono and Paniconi \citep{oono1998steady}.
In what follows, we are going to find the integrating factor and the
related potential. As we will see, they define the nonequilibrium
temperature and $S^{*}$, which may be called a nonequilibrium thermodynamic
entropy.

Before proceeding further, it is worth giving several comments. First,
for constant $N,$ four parameters determine the state of the system,
$T_{1},T_{2},A,L$. So the Pfaff form for the heat (\ref{eq:ex heat differential 2})
may be written in the space of these parameters in terms of $dT_{1}$,
$dT_{2}$ , $dA$, and $dL$. Second, because the pressure in the
system is homogeneous, we can write the expression for elementary
work, $\mkern3mu \mathchar'26\mkern-12mu dW=-pAdL-pLdA$, which we
shortly write in terms of the volume of the system, $\mkern3mu \mathchar'26\mkern-12mu dW=-pdV$.
Third, once the integrating factor and the corresponding potential
are found, it is straightforward to represent them in another set
of variables of states. It is easy to check, that the integrating
factor in variables $X$, denoted by $\lambda_{x}\left(X\right)$,
after changing the variables of states to $Y$ given by $Y\left(X\right)$,
transforms to $\lambda_{y}\left(Y\right)=\lambda_{x}\left(X\left(Y\right)\right)$.
Similar holds for the potential corresponding to the integrating factor.
We work in variables $U,V,T_{2}/T_{1}$. In these variables, the heat
differential (\ref{eq:ex heat differential 2}) is given by, 
\begin{equation}
\mkern3mu \mathchar'26\mkern-12mu dQ=dU+\frac{2}{3}\frac{U}{V}dV+0\cdot d\frac{T_{2}}{T_{1}},\label{eq:heat by U and V}
\end{equation}
where we explicitly wrote the vanishing third term to remind that
the form is in three-dimensional space, $U,V,T_{2}/T_{1}$ and used
expression (\ref{eq:pressure U and V}) for pressure.

We observe that the heat differential (\ref{eq:heat by U and V})
in variables $U,V,T_{2}/T_{1}$ is identical to the heat differential
for an ideal gas in equilibrium thermodynamics \citep{Thermodynamics_and_an_Introduction_to_Thermostatistics_2ed_H_Callen}.
This is a consequence of the fact that both in equilibrium thermodynamics
and in the nonequilibrium stationary state considered here, the energy
is exchanged in two same ways (heat and mechanical work) and that
the relationship between pressure and internal energy for equilibrium
ideal gas, $p=2U/3V$, is identical to formula (\ref{eq:pressure U and V}).
Therefore, the heat differential has an integrating factor $T^{*}\left(U,V,T_{2}/T_{1}\right)$
and the corresponding potential, $S^{*}\left(U,V,T_{2}/T_{1}\right)$,

\begin{equation}
\mkern3mu \mathchar'26\mkern-12mu dQ=T^{*}dS^{*}.\label{eq:integrating factor}
\end{equation}
The integrating factor and the potential are not unique. To find the
integrating factor we observe that formula (\ref{eq:heat by U and V})
is the thermodynamics first law for a monoatomic ideal gas in equilibrium
thermodynamics. In this case, the integrating factor is the temperature
of the system, which for an ideal gas is given by the formula $T=2U/3Nk_{B}$.
For the nonequilibrium case considered here, we introduce a similar
expression, so the integrating factor is given by 
\begin{equation}
T^{*}=\frac{2U}{3Nk_{B}}.\label{eq:tstar}
\end{equation}
The potential corresponding to the above integrating factor is $S^{*}$.
As follows from (\ref{eq:heat by U and V}), (\ref{eq:integrating factor})
and (\ref{eq:tstar}), the differential of $S^{*}$ is given by, 
\[
dS^{*}=\frac{3Nk_{B}}{2U}dU+\frac{Nk_{B}}{V}dV.
\]
$S^{*}$ is thus given by the following formula, $S^{*}\left(U,A,L,T_{2}/T_{1}\right)=Nk_{B}\log\left(U^{3/2}V\right)+S_{0},$
where $S_{0}$ is a numeric constant. However, it may depend on parameters
of the system which are not treated as the variables of state, including
the number of particles $N$, which we set to be constant in the above
reasoning. We determine $S_{0}$ constant by the condition that $S^{*}$
for $T_{2}=T_{1}$ gives the equilibrium expression \citep{Thermodynamics_and_an_Introduction_to_Thermostatistics_2ed_H_Callen}.
Therefore we get, 
\begin{equation}
S^{*}\left(U,V,T_{2}/T_{1}\right)=Nk_{B}\left\{ \frac{5}{2}+\frac{3}{2}\log\left[\frac{2}{3}\frac{\varphi_{0}U}{N}\left(\frac{V}{N}\right)^{2/3}\right]\right\} ,\label{eq:noneq entropy}
\end{equation}
where $\varphi_{0}$ is a constant that does not depend on any control
parameter. The above fundamental relation has proper partial derivatives,
\begin{align}
\left(\frac{\partial S^{*}}{\partial U}\right)_{V,N} & =\frac{1}{T^{*}},\label{eq:temp st by partial}\\
\left(\frac{\partial S^{*}}{\partial V}\right)_{U,N} & =\frac{p}{T^{*}}.\nonumber 
\end{align}

As a potential of the heat differential, the above $S^{*}$ determines
stationary-state adiabats \citep{oono1998steady}. They are different
from adiabats defined in equilibrium. Because $S^{*}$ does not depend
on the temperature ratio, we see that the change of $T_{2}/T_{1}$
(keeping $U,A,L$, and $N$ constant) changes the temperature profile
in the system. It also changes the heat flowing through the system.
But it does not trigger the exchange of the net heat. As we show below,
$T_{2}/T_{1}$ is a parameter that controls the entropy production
in the system.

There is a natural question about the relation between $S^{*}$ and
the total entropy of the system, $S_{\text{tot}}=A\int dz\,s\left(z\right)$,
where $s\left(z\right)$ is the volumetric entropy density 
\begin{equation}
s\left(z\right)=n\left(z\right)k_{B}\left\{ \frac{5}{2}+\frac{3}{2}\log\left[\varphi_{0}k_{B}T\left(z\right)\left[n\left(z\right)\right]^{-2/3}\right]\right\} ,\label{eq:s dens}
\end{equation}
as given by local equilibrium assumption within irreversible thermodynamics
\citep{Groot_Mazur_Non-equilibrium_thermodynamics,Thermodynamics_and_an_Introduction_to_Thermostatistics_2ed_H_Callen}.
With the use of the linear temperature profile and density determined
above, we obtain, 
\begin{align}
S_{\text{tot}}\left(U,V,\frac{T_{2}}{T_{1}}\right) & =S^{*}\left(U,V\right)+\Delta S\left(U,V,\frac{T_{2}}{T_{1}}\right),\label{eq:s star and total}\\
\Delta S\left(U,V,T_{2}/T_{1}\right) & =Nk_{B}\log\left[\left(\frac{T_{2}}{T_{1}}\right)^{5/4}\left(\frac{\log\frac{T_{2}}{T_{1}}}{\frac{T_{2}}{T_{1}}-1}\right)^{5/2}\right].\nonumber 
\end{align}
The above expression is symmetric with respect to the interchange
of $T_{1}$ and $T_{2}$. Only $S^{*}$ contains information about
heat absorbed/released in the system (see Eqs (\ref{eq:ex heat differential 2},\ref{eq:integrating factor}))
on top of the dissipative background (temperature profile). $\Delta S$,
on the other hand, controls the dissipative background since it depends
on the entropy production given by \citep{Groot_Mazur_Non-equilibrium_thermodynamics}
$\sigma=-A\int_{0}^{L}dz\,\kappa\nabla T\left(z\right)\cdot\nabla\frac{1}{T\left(z\right)}=\frac{A\kappa}{L}\left(\frac{T_{2}}{T_{1}}+\frac{T_{1}}{T_{2}}-2\right)$.
The difference between the total entropy and $S^{*}$ vanishes, $\Delta S\left(U,A,L,T_{2}/T_{1}\right)\to0$,
when the system approaches the equilibrium state, $T_{2}/T_{1}\to1$.
Therefore, the $S^{*}$ becomes in this limit the equilibrium entropy.

The relation between the equilibrium entropy and $S^{*}$ also sheds
light on the role of $T_{2}/T_{1}$ parameter. For an ``adiabatically''
insulated system determined by condition $S^{*}=const$, the parameter
$T_{2}/T_{1}$ changes the total entropy of the system. The change
of the total entropy of the system is associated with the local heat
transfer and work between different subparts of the nonequilibrium
system, even if no work is performed on the system and no net heat
enters it.

\section{Zeroth and second law for nonequilibrium stationary states}

In the above, we showed that a net heat potential, nonequilibrium
entropy $S^{*}$, exists for the system without a separating wall.
Here we consider the existence of the potential in the context of
the system from Fig. 1 with an internal wall. It is a diathermic wall
that separates the gas. We assume that the wall is at position $l_{w}$
and there are $N_{1}$ particles to the left and $N_{2}=N-N_{1}$
particles to the right of the wall. An external force, $F_{w},$ can
move the wall, and some work is related when the wall moves. As before,
the system is described by thermohydrodynamic equations, this time
with additional boundary conditions on the surface of the separating
wall. At the stationary state, the pressure is homogeneous in each
subsystem, but they may be different due to the action of the force
on the wall. In the stationary state $F_{w}=-A\left(p_{2}-p_{1}\right)$.
We assume that the wall is diathermic, so the temperature profile
is the same as for the system without the wall. The temperature profile
does not depend on the action of the force on the wall. We notice
that each subsystem looks like the system without the wall, although
with different parameters, so we can use formulas for the system without
the wall to describe the system with the wall. 

We describe the system's energy with the wall using the state variables
for each subsystem, $U\left(S_{1}^{*},V_{1},N_{1},S_{2}^{*},V_{2},N_{2}\right)=U_{1}\left(S_{1}^{*},V_{1},N_{1}\right)+U_{2}\left(S_{2}^{*},V_{2},N_{2}\right)$.
The additivity of the energy is inscribed in the used thermohydrodynamic
equations. But the additivity of entropy is a postulate of equilibrium
thermodynamics. The nonequilibrium entropy $S^{*}$ is not additive
for a nonequilibrium system with heat flow. If the entropy was additive,
then expression $\Gamma\left(U_{1},U_{2},V_{1},V_{2},N_{1},N_{2}\right)\equiv S^{*}\left(U_{1}+U_{2},V_{1}+V_{2},N_{1}+N_{2}\right)-S^{*}\left(U_{1},V_{1},N_{1}\right)-S^{*}\left(U_{2},V_{2},N_{2}\right)$
would identically be zero. Here 
\begin{equation}
S^{*}\left(U,V,N\right)\equiv Nk_{B}\left\{ \frac{5}{2}+\frac{3}{2}\log\left[\frac{2}{3}\frac{\varphi_{0}U}{N}\left(\frac{V}{N}\right)^{2/3}\right]\right\} \label{eq:entropy UVN}
\end{equation}
in agreement with Eq. (\ref{eq:noneq entropy}). It is cumbersome
to show by explicit calculations that $\Gamma$ does not vanish. Instead,
we calculate the following expression, 
\[
\frac{\partial}{\partial U_{2}}\left[\left(U_{1}+U_{2}\right)U_{1}\left(\frac{\partial\Gamma}{\partial U_{1}}\right)\right]=-\frac{3}{2}N_{1}k_{B},
\]
which proves, that $\Gamma$ cannot vanish identically. Therefore
$S^{*}\neq S_{1}^{*}+S_{2}^{*}$ for most states, and the nonequilibrium
entropy is not additive.

One can wonder why the nonequilibrium entropy is not additive. Yet
the nonequilibrium entropy of each subsystem is given by the same
formula for the equilibrium situation, i.e. Eq. (\ref{eq:entropy UVN}).
In equilibrium, the zeroth law of thermodynamics would allow us to
introduce the additive entropy of the whole system when the entropy
of two subsystems is known. The total heat differential is given by,
$\mathchar'26\mkern-12mu dQ=\mathchar'26\mkern-12mu dQ_{1}+\mathchar'26\mkern-12mu dQ_{2}=T_{1}dS_{1}+T_{2}dS_{2}$.
This heat differential in the space of parameters $U_{1},V_{1},N_{1},U_{2},V_{2},N_{2}$
has no integrating factor. But the zeroth law imposes the condition
of equal temperatures, $T_{1}=T_{2}\equiv T$, simplifying the heat
differential to, $\mathchar'26\mkern-12mu dQ=T\left(dS_{1}+dS_{2}\right)$.
We see that the function $S$ defined by $S\equiv S_{1}+S_{2}$ is
a potential of heat with the temperature $T$ as the integrating factor.
That is how the equilibrium zeroth law leads to the additivity of
entropy.

For a nonequilibrium system, there is no equality of subsystems' temperatures.
The equilibrium zeroth law of thermodynamics is broken. However, let's
introduce the following condition called the \textquotedbl zeroth
law of global stationary thermodynamics\textquotedbl{} for the ideal
gas with a heat flow in the following form,

\begin{equation}
T_{2}^{*}=rT_{1}^{*}.\label{eq:zeroth law}
\end{equation}
with a constant parameter $r$. With the above zeroth law condition,
the net heat differential is given by $\mathchar'26\mkern-12mu dQ=T_{1}^{*}dS_{1}^{*}+rT_{1}^{*}dS_{2}^{*}.$
It appears that it has an integrating factor that is easy to guess.
Defining a function 
\begin{equation}
S_{12}^{*}\equiv S_{1}^{*}+rS_{2}^{*}\label{eq:none S for two subsystems}
\end{equation}
allows us to represent the above heat differential by 
\[
\mathchar'26\mkern-12mu dQ=T_{1}^{*}dS_{12}^{*}.
\]
For every given nonequilibrium temperature ratio $r$, which appears
in the zeroth law condition (\ref{eq:zeroth law}), the above nonequilibrium
entropy $S_{12}^{*}$ splits the space of thermodynamic parameters
$U_{1},V_{1},N_{1},U_{2},V_{2},N_{2}$ on adiabatically insulated
subspaces parametrized by $S_{12}^{*}$.

We are now in a position to verify whether the nonequilibrium entropy
$S_{12}^{*}$ can be used to generalize the equilibrium minimum energy
principle to the case with the heat flow. We check whether the minimization
of the energy for constant nonequilibrium entropy $S_{12}^{*}$ leads
to the proper position of a movable wall. The verification requires
assuming a constant number of particles, $N_{1},N_{2}$, total volume,
$V=V_{1}+V_{2},$ and nonequilibrium entropy $S_{12}^{*}$ given by
(\ref{eq:none S for two subsystems}) and calculate the minimum of
the total energy, 
\begin{align}
 & U_{\text{tot}}\left(S_{1}^{*},V_{1}\right)\equiv\nonumber \\
 & U_{1}\left(S_{1}^{*},V_{1},N_{1}\right)+U_{2}\left(\frac{1}{r}\left(S_{12}^{*}-S_{1}^{*}\right),V-V_{1},N_{2}\right),\label{eq:energy to minimize}
\end{align}
where 
\begin{equation}
U_{1}\left(S_{1}^{*},V_{1},N_{1}\right)=\frac{3}{2\varphi_{0}}N_{1}\left(\frac{V_{1}}{N_{1}}\right)^{-\frac{2}{3}}\exp\left[\frac{S_{1}^{*}}{N_{1}k_{B}}-\frac{5}{3}\right]\label{eq:U by SVN}
\end{equation}
and is obtained from Eq. (\ref{eq:entropy UVN}). $U_{2}$ is given
by a similar formula. The total energy in Eq. (\ref{eq:energy to minimize})
have two independent parameters, $S_{1}^{*}$ and $V_{1}$. The minimum
of the above energy requires the vanishing of the derivatives over
the two independent parameters, which gives 
\begin{align}
T_{1}^{*}-\frac{1}{r}T_{2}^{*} & =0,\nonumber \\
p_{1}-p_{2} & =0.\label{eq:extr cond}
\end{align}
The above two equations determine the two independent parameters.
From thermohydrodynamics, we know that the equality of pressures is
the proper condition for the position of the movable wall. Equivalently,
the vanishing of the derivatives of $U_{\text{tot}}\left(S_{1}^{*},V_{1}\right)$
leads to,
\begin{align}
V_{1} & =\frac{V}{N_{2}\left(\frac{r}{N_{1}}+\frac{1}{N_{2}}\right)},\nonumber \\
S_{1}^{*} & =\frac{N_{1}}{N_{1}+rN_{2}}S_{12}^{*}-\frac{5}{2}k_{B}\frac{rN_{1}N_{2}}{N_{1}+rN_{2}}\log r.\label{eq:v1 s1 minimum}
\end{align}
It follows that for positive $r,$ arbitrary entropy $S_{12}^{*}$,
and fixed $N_{1},N_{2}$ and $V$, there is at most one point in space
$S_{1}^{*}$,$V_{1}$ with vanishing derivatives. Because of the simple
form of $U_{\text{tot}}$, it is easy to show that $S_{1}^{*}$ and
$V_{1}$ given by Eqs. (\ref{eq:v1 s1 minimum}) are at the global
minimum. It proves that the equilibrium minimum energy principle generalizes
to the case with heat flow.

The above application of the ``second law'' requires constant $N_{1},N_{2},V=V_{1}+V_{2}$,
the nonequilibrium temperature ratio $r$ and the nonequilibrium entropy
$S_{12}^{*}$. To realize it experimentally, one has to know the boundary
temperatures $T_{1}$ and $T_{2}$ as a function of $S_{12}^{*},r,V_{1},V,N_{1}$
and $N_{2}$. Utilizing Eqs. (\ref{eq:pressure U and V}), (\ref{eq:p by T1 T2})
and (\ref{eq:tstar}) leads to the following nonequilibrium temperatures
for both subsystems, 
\begin{align*}
T_{1}^{*} & =\frac{\frac{V_{1}}{V}\left(T_{2}-T_{1}\right)}{\log\left(\frac{T_{1}+\frac{V_{1}}{V}\left(T_{2}-T_{1}\right)}{T_{1}}\right)},\\
T_{2}^{*} & =\frac{\left(1-\frac{V_{1}}{V}\right)\left(T_{2}-T_{1}\right)}{\log\left[\frac{T_{2}}{T_{1}+\frac{V_{1}}{V}\left(T_{2}-T_{1}\right)}\right]}.
\end{align*}
Using the above expressions in zeroth law condition (\ref{eq:zeroth law}),
we obtain 
\[
\left(1-\frac{V_{1}}{V}\right)\log\left[1+\frac{V_{1}}{V}\left(\tau-1\right)\right]=r\frac{V_{1}}{V}\log\left[\frac{\tau}{1+\frac{V_{1}}{V}\left(\tau-1\right)}\right],
\]
with the boundary temperatures ratio $\tau=T_{2}/T_{1}$. The above
relation implicitly determines the boundary temperature ratio as a
function of $V_{1}/V$, i.e. $\tau=\tau\left(V_{1}/V\right).$ Because
this relation is implicit, it is impossible to determine $T_{1}\left(S_{12}^{*},r,V_{1},V,N_{1},N_{2}\right)$
explicitly. But having $\tau\left(V_{1}/V\right)$, we may use expressions
(\ref{eq:entropy UVN}), (\ref{eq:zeroth law}), (\ref{eq:none S for two subsystems})
and (\ref{eq:temp st by partial}) to determine $T_{1}\left(S_{12}^{*},r,V_{1},V,N_{1},N_{2}\right)$.

It is straightforward to generalize the above conclusions for any
system shape and temperature profile. The reason for that is the fact
that a particular form of the temperature profile does not play a
role in the above calculations - the existence of the global steady
state thermodynamics follows in the considered case from the fact
that pressure is constant and it is a function of energy and volume,
here $p=2U/3V$. The ideal gas in the box volumetrically heated and
separated by a movable wall considered by Zhang et al. \citep{Zhang2021continuous}
also has these properties. Therefore, the steady state global thermodynamics
formulated here also holds for Zhang et al. \citep{Zhang2021continuous}
system, describing the continuous phase transition they consider.
We describe Zhang et al. case in the next section.

\section{Ideal gas under volumetric heat supply}

For the volumetrically heated gas the heat rate in Eq. (\ref{eq:heat rate})
does not vanish in stationary state, $q\left(t\right)\neq0$. Zhang
{et al.}~\citep{Zhang2021continuous} consider ideal gas with uniform
volumetric heating $\lambda$. In this case, in the energy balance
(\ref{eq:energy balance}) there appears the source term $\lambda$
on the right-hand side. As a consequence, Eq. (\ref{eq:du by time integral})
is modified in the following way, $dU=\int_{t_{i}}^{t_{f}}dt\,q\left(t\right)-\int_{t_{i}}^{t_{f}}dt\,p\left(t\right)\frac{dV\left(t\right)}{dt}+\lambda\int_{t_{i}}^{t_{f}}dt\,V\left(t\right),$
where, as before, we take the dominant term in, $\int_{t_{i}}^{t_{f}}dt\,p\left(t\right)\frac{dV\left(t\right)}{dt}\approx pdV$,
and obtain, 
\[
dU=\int_{t_{i}}^{t_{f}}dt\,q\left(t\right)-pdV+\lambda\int_{t_{i}}^{t_{f}}dt\,V\left(t\right).
\]
Using the above and defining 
\begin{equation}
\mkern3mu \mathchar'26\mkern-12mu dQ=\int_{t_{i}}^{t_{f}}dt\,q\left(t\right)+\lambda\int_{t_{i}}^{t_{f}}dt\,V\left(t\right)\label{eq:excess heat}
\end{equation}
we obtain Eq. (\ref{eq:energy balance dU}). It is worth commenting
on the fact that for $\lambda\neq0$, both the net heat $\int_{t_{i}}^{t_{f}}dt\,q\left(t\right)$
and the second term in the right-hand side of the above expression
are infinite in the limit of long transition between two neighboring
stationary states. However, their sum is finite. In this system, the
heat constantly flows out of the system, $q\left(t\right)\neq0$.
The outflow is balanced by the generation of heat within the system
as given by the heat generation rate, $\lambda V\left(t\right)$.
In Oono and Paniconi~\citep{oono1998steady} terminology, $\mkern3mu \mathchar'26\mkern-12mu dQ$
is the excess heat, and $\lambda V\left(t\right)$ is the house-keeping
heat rate. The above equation is interpreted as the ``renormalization''
of the heat rate to obtain the excess heat \citep{oono1998steady}.

Zhang et al. \citep{Zhang2021continuous} considered an ideal gas
between two parallel walls of area $A$ located at $z=-L$ and $z=L$.
The system is translationally invariant in the $x$ and $y$ directions.
The walls are kept at a fixed temperature $T_{0}$ and the energy
is supplied into the system's volume in the form of heat with the
flux $J$; the supplied energy per unit time and unit volume $V=2AL$
is $\lambda=J/V$. The steady state temperature profile can be obtained
from the local continuity equation of energy 
\begin{equation}
-\kappa\dfrac{\partial^{2}}{\partial z^{2}}T(z)=\lambda
\end{equation}
with the boundary conditions $T(0)=T(L)=T_{0}$, giving 
\begin{equation}
T(z)=-\dfrac{\lambda}{2\kappa}z^{2}+\dfrac{\lambda}{2\kappa}L^{2}+T_{0}.\label{eqn:app-T}
\end{equation}
At the steady state the pressure $p$ and hence also the energy density
$\epsilon$ are constant. With the use of the equation of state, this
determines the density profile 
\begin{equation}
n\left(z\right)=\frac{p}{k_{B}T\left(z\right)}=\frac{2}{3}\frac{\epsilon}{k_{B}T\left(z\right)}.\label{eqn:density}
\end{equation}
Using 
\begin{equation}
N=A\int_{-L}^{L}dz\,n\left(z\right),
\end{equation}
for a given number of particles $N$, the energy $U=\epsilon V$ is
obtained as 
\begin{equation}
U=%U_{eq}f(\lambda\cdot\ddfrac{L^{2}}{\kappaT_{0}})=
\dfrac{3}{2}Nk_{B}T_{0}f(\lambda\cdot\dfrac{L^{2}}{\kappa T_{0}}),\label{eqn:app-u}
\end{equation}
where a dimensionless function $f$ is given by 
\begin{equation}
f(x)\equiv\sqrt{x(x+2)}/(2\tanh^{-1}\sqrt{x/(x+2)}).\label{eq:func}
\end{equation}
Using the volumetric entropy density of an ideal gas given by Eq.~(\ref{eq:s dens}),
we obtain 
\[
S_{\text{tot}}\left(U,A,L,N,\lambda\right)=S^{*}\left(U,V,N\right)+\Delta S\left(U,A,L,N,\lambda\right),
\]
where $S^{*}\left(U,V,N\right)$ has a form given by the RHS of Eq.~(\ref{eq:noneq entropy})
and 
\begin{align}
\Delta S\left(U,A,L,N,\lambda\right) & =-(5/2)Nk_{B}\log\left[f(\lambda\cdot\dfrac{L^{2}}{\kappa T_{0}})\right]\nonumber \\
 & +(5/3)\epsilon A\int_{0}^{L}dz\,\frac{\log\left[T\left(z\right)/T_{0}\right]}{T\left(z\right)}.
\end{align}
In the above formula, $T_{0}\left(U,A,L,N,\lambda\right)$ is implicitly
given by Eq.~(\ref{eqn:app-u}). The temperature profile is a quadratic
function of the distance $z$, therefore the integral in the above
equation cannot be expressed in terms of elementary functions. Nevertheless,
as for the case discussed in the previous section, the nonequilibrium
entropy differs from the total entropy in the system.

Now we introduce a movable adiabatic wall parallel to the bounding
walls at $z=z_{w}$. At equilibrium, the wall is located precisely
in the middle of the system $z_{w}=0$. As shown in Ref.~\citep{Zhang2021continuous},
for small heat fluxes, the position of the wall at $z_{w}=0$ is stable.
Above a critical flux, the wall moves towards one of the bounding
surfaces. Let us consider the second law of nonequilibrium thermodynamics
discussed above for this system. Integration of the equation of state
for the ideal gas (Eq.~(\ref{eqn:density})) in each subsystem 1
and 2 leads to 
\begin{equation}
p_{1,2}=\frac{2}{3}\frac{U_{1,2}}{V_{1,2}},
\end{equation}
and for each subsystem Eq.~(\ref{eq:heat by U and V}) is satisfied
(with the replacement $T_{2}/T_{1}\rightarrow\lambda$). This implies
the existence of nonequilibrium entropy and nonequilibrium temperature
given by formulas (\ref{eq:tstar}) and (\ref{eq:noneq entropy}).

Consequently, the reasoning leading to the minimum principle is the
same. The only difference is that instead of two temperatures $T_{1},T_{2}$,
we have here a single temperature $T_{0}$ of the confining walls
and the volumetric heating rate $\lambda$. The nonequilibrium temperatures
in this case are given by: 
\begin{align}
T_{1}^{*} & =T_{0}f\left(\frac{\lambda L_{1}^{2}}{\kappa T_{0}}\right)\nonumber \\
T_{2}^{*} & =T_{0}f\left(\frac{\lambda L_{2}^{2}}{\kappa T_{0}}\right),\label{eqn:app-T1,2}
\end{align}
where $L_{1}=L+z_{w}$ and $L_{2}=L-z_{w}$ and the function $f$
given by (\ref{eq:func}). In this case, the minimization of energy
(\ref{eq:energy to minimize}) also leads to the equality of pressures.
It proves that ideal gas with volumetric heating can also be described
with three laws of global stationary thermodynamics introduced in
the previous section.

The minimization principle introduced in the previous section also
leads to a single global minimum with the zero law condition and equality
of pressures given by (\ref{eq:extr cond}). Interestingly, as discussed
by Zhang et al., this system exhibits a continuous phase transition
from a one stable steady state with the wall in the middle of the
system to the two stable stationary states with the mirror symmetry.
On the other hand, the minimization procedure leads to only one stable
state. However, the zeroth law condition, $r=T_{2}^{*}/T_{1}^{*}$,
breaks the symmetry. Application of formulas (\ref{eqn:app-T1,2})
in the zeroth law condition, $r=T_{2}^{*}/T_{1}^{*}$, leads to the
conclusion that for $V_{1}<V_{2}$, the parameter $r$ takes only
values $r>1$. Whereas for $V_{1}>V_{2}$, the temperature ratio satisfies
$0<r<1$. Therefore, setting $r$ limits the motion of the wall to
half of the system. The second stable minimum is obtained by replacing
$r$ with $1/r$.

\section{Conclusions}

It is straightforward to generalize the above conclusions for the
situation with the temperature-dependent heat conductivity, $\kappa\left(T\right),$
which is beyond the scope of linear irreversible thermodynamics. The
temperature-dependent heat conductivity modifies the Fourier law (\ref{eq:fourier law})
and the temperature profile. However, it does not affect the relation,
$p=2U/3V$. Therefore, the relation (\ref{eq:heat by U and V}) holds
and it is possible to repeat the reasoning presented above without
any changes and obtain $S^{*}$ given by Eq. (\ref{eq:noneq entropy})
and the nonequilibrium temperature (\ref{eq:tstar}). It is worth
noting that the assumption of local equilibrium for the ideal gas
is valid as long as the temperature gradient is sufficiently small,
$l_{fp}\left|\nabla T\right|/T\ll1,$ for the mean free path of the
molecules, $l_{fp}$ \citep{Nonequilibrium_thermodynamics_and_its_statistical_foundations_H_J_Kreuzer}.
At the pressure of $1$ bar at room temperature, the mean free path
is of the order of $l_{fp}\approx100nm$. The local equilibrium is
satisfied as long as the temperature gradient is lower than ten million
Kelvins per centimeter.

We draw several conclusions from the rigorous calculations performed
above within irreversible thermodynamics. Even for the system with
heat flow, which is far from equilibrium (significant temperature
difference), global steady state thermodynamics exist. This is for
ideal gas closed in a vessel of any shape and does not depend on the
mode of the heat transfer (heat flows through the system or the system
is heated volumetrically). The considered examples also show that
$S^{*}$ that governs the net heat is independent of the entropy production,
in agreement with Eq. (\ref{eq:s star and total}).

At least since the works of Prigogine, scientists have tried to formulate
a thermodynamic-like description of nonequilibrium systems. Here we
show that it exists for a stationary ideal gas with heat flow. The
question remains open for interacting systems and systems with kinetic
energy. From the perspective of future efforts, the case of ideal
gas considered here shows that the local entropy integrated over volume
is not a quantity that determines the heat in the system - a possibility
discussed recently \cite{nakagawa2019global}. Moreover, the considered
case shows that nonequilibrium entropy, defined as heat potential,
is not additive.

\section*{Acknowledgements}

P.J.Z. would like to acknowledge the support of a project that has
received funding from the European Union\textquoteright s Horizon
2020 research and innovation program under the Marie Sk\l odowska-Curie
Grant Agreement No. 847413 and was a part of an international cofinanced
project founded from the program of the Minister of Science and Higher
Education entitled \textquotedblleft PMW\textquotedblright{} in the
years 2020--2024; Agreement No. 5005/H2020-MSCA-COFUND/2019/2.

\bibliographystyle{unsrt}
%\bibliography{/home/karol/Desktop/nauka/artykuly/bib_file/baza_artukolow,/home/karol/Desktop/nauka/artykuly/bib_file/ksiazki}

\end{document}